\documentclass[preprint,showpacs,preprintnumbers,amsmath,amssymb]{revtex4}

\usepackage{graphicx}
\usepackage{dcolumn}
\usepackage{bm}
\usepackage{bbm}

\begin{document}


\title{Atomic Structure of Steps on $\bf{180^{\circ}}$ Ferroelectric \\ Domain Walls in PbTiO$_3$}

\author{Arzhang Angoshtari}
\author{Arash Yavari}
 \email{arash.yavari@ce.gatech.edu}
\affiliation{ School of Civil and Environmental Engineering, Georgia
Institute of Technology, Atlanta, GA 30332. }

\date{\today}



\begin{abstract}
Using the method of anharmonic lattice statics, we calculate the equilibrium structure of steps on $180^{\circ}$ ferroelectric domain
walls (DW) in PbTiO$_3$. We consider three different types of steps: i) Ti-Ti step that joins a Ti-centered DW to a Ti-centered DW, (ii) Pb-Pb step that joins a Pb-centered DW to a Pb-centered DW, and (iii) Pb-Ti step that joins a Pb-centered DW to a Ti-centered DW. We show that atomic distortions due to these steps broaden a DW but are localized, i.e., they are confined to regions with dimensions of a few lattice spacings. We see that a step locally thickens the domain wall; the defective domain wall is two to three times thicker than the perfect domain wall depending on the step type. We also observe that steps distort the polarization distribution in a mixed Bloch-N\'{e}el like way; polarization rotates out of the domain wall plane near the steps. Our calculations show that Pb-Pb steps have the lowest static energy.
\end{abstract}


\maketitle

\section{Introduction}

Ferroelectric materials are an important subclass of polar materials
due to their wide range of applications in ultrasound imaging,
microelectromechanical systems, high strain actuators,
electro-optical systems, photothermal imaging, and high density
storage devices \cite{Kalinin2010}. It is known that some important
properties of ferroelectric materials are due to the presence of
domain walls, which are two-dimensional defects that separate
regions with uniform polarization \cite{DawberRabeScott2005}. This
explains the importance of a detailed study of the properties of
the domain walls.

\begin{figure*}[t]
\centerline{\hbox{\includegraphics[scale=0.65,angle=0]{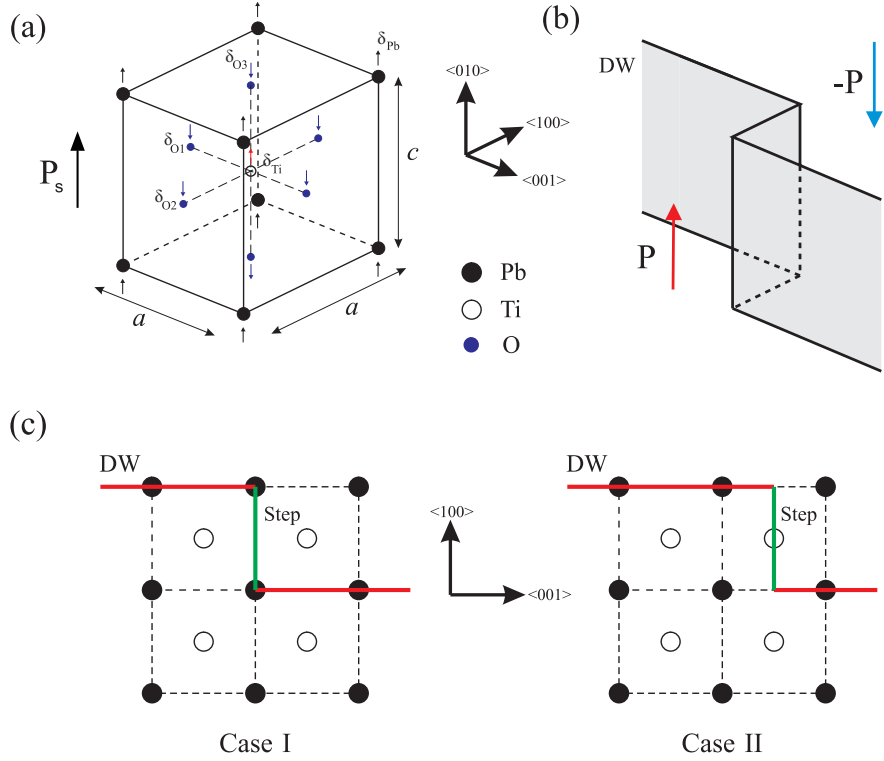}
\vspace*{-0.2in}
                 }}
\caption{ (a) The relaxed configuration of the unit cell of tetragonal
PbTiO$_3$. $a$ and $c$ are the tetragonal lattice parameters. Note
that O1, O2, and O3 refer to oxygen atoms located on $(001)$,
$(100)$, and $(010)$-planes, respectively. $\delta$ denotes the
y-displacements of the atoms from the centerosymmetric position and
arrows near each atom denote the direction of these displacements.
(b) Schematic profile of polarization close to a step. (c) Two
different possibilities for a Pb-Pb step. } \label{UC}
\end{figure*}

From both experimental and theoretical studies, it is observed that
the thickness of domain walls can vary from a few angstroms
\cite{MeyerVanderbilt2001,Padilla1996,YaOrBh2006b,AngYa2010,HlinkaMarton2006}
to a few micrometers \cite{Iwata2003,Lehnen2000}. It has been suggested
that this wide scatter in the domain wall thickness is due to the
presence of point defects \cite{Shilo2004,Lee2005,AngYavari2010}.
Another important property of domain walls is the behavior of the
polarization profile near the domain wall. It is well known that
$180^{\circ}$ domain walls have an Ising-like nature. Using
Monte-Carlo simulation, Padilla \textit{et al.} \cite{Padilla1996}
showed the predominant Ising-like character of $180^{\circ}$ domain
walls in tetragonal BaTiO$_3$ along the tetragonal axis. In
$180^{\circ}$ domain walls, polarization vector can either rotate in
a plane parallel to the domain wall (Bloch type) or normal to the
domain wall (N\'{e}el type) \cite{Lee2009}. Subsequent works on the
domain walls showed that domain walls can have mixed characters.
Using density functional theory, Lee \textit{et al.} \cite{Lee2009}
showed that while $180^{\circ}$ domain walls in PbTiO$_3$ are
predominantly Ising-like, they have some N\'{e}el characters as
well. Having the domain walls parallel to the $(100)$-plane, we know
that polarization is mainly along the $\langle010\rangle$-direction (see Fig.
\ref{UC}). As Lee \textit{et al.} \cite{Lee2009} showed close to the
domain wall polarization has normal components (normal to the
domain wall) with magnitudes in the order of 1-2 percent of the bulk
polarization. Angoshtari and Yavari \cite{AngYa2010} observed a
similar behavior at finite temperatures for perfect $180^{\circ}$
domain walls. They saw normal components in the order of 2 percent
of the bulk polarization in their finite-temperature structure
calculations. Recently, first-principle-based simulations have led
to the prediction of vortex type polarization distribution in
zero-dimensional ferroelectric nanodots
\cite{NaumovBellaicheFu2004,Prosandeev2008}.

It is believed that steps have an important role in domain wall
motion. Nettleton \cite{Nettleton1967} proposed a model for sidewise
displacement of a $180^{\circ}$ domain wall in a single crystal
barium titanate and suggested that the formation of an irregular
pattern of steps of varying shapes and sizes results in the motion
of the domain wall and the speed of the domain wall motion is
determined by the rate of formation and disappearance of these
steps. Shur \textit{et al.} \cite{Shur1990} considered steps on
$180^{\circ}$ domain walls and proposed a mechanism for domain wall
motion in weak and strong fields. Shin \textit{et al.}
\cite{ShinGrinberg2007} used atomistic molecular dynamics and
coarse-grained Monte Carlo simulations to analyze the nucleation and
growth mechanism of domain walls in PbTiO$_3$ and BaTiO$_3$.

In this work we investigate the effect of steps, which are one-dimensional defects, on $180^{\circ}$ domain walls
parallel to $(100)$-planes in PbTiO$_3$ using the anharmonic lattice
statics method. We consider Ti-Ti steps that join a Ti-centered DW
to another Ti-centered DW, Pb-Pb steps that join a Pb-centered DW to
another Pb-centered DW, and Pb-Ti steps that join a Pb-centered DW
to a Ti-centered DW. As the initial configuration, we start from the
atomic configuration of perfect $180^{\circ}$ domain walls and then
relax the structure iteratively to obtain the optimized atomic
configuration.

This paper is organized as follows. In \S\ref{sec:Geometry}, we
explain the initial geometry of steps that we analyzed throughout
this work. In \S\ref{sec:Calculation}, we discuss the method of
anharmonic lattice statics and the shell potential for PbTiO$_3$
that we used in our calculations. We present our numerical results
in \S\ref{sec:Results}. The paper ends with concluding remarks in
\S\ref{sec:Concluding}.

\begin{figure*}[t]
\centerline{\hbox{\includegraphics[scale=1.05,angle=0]{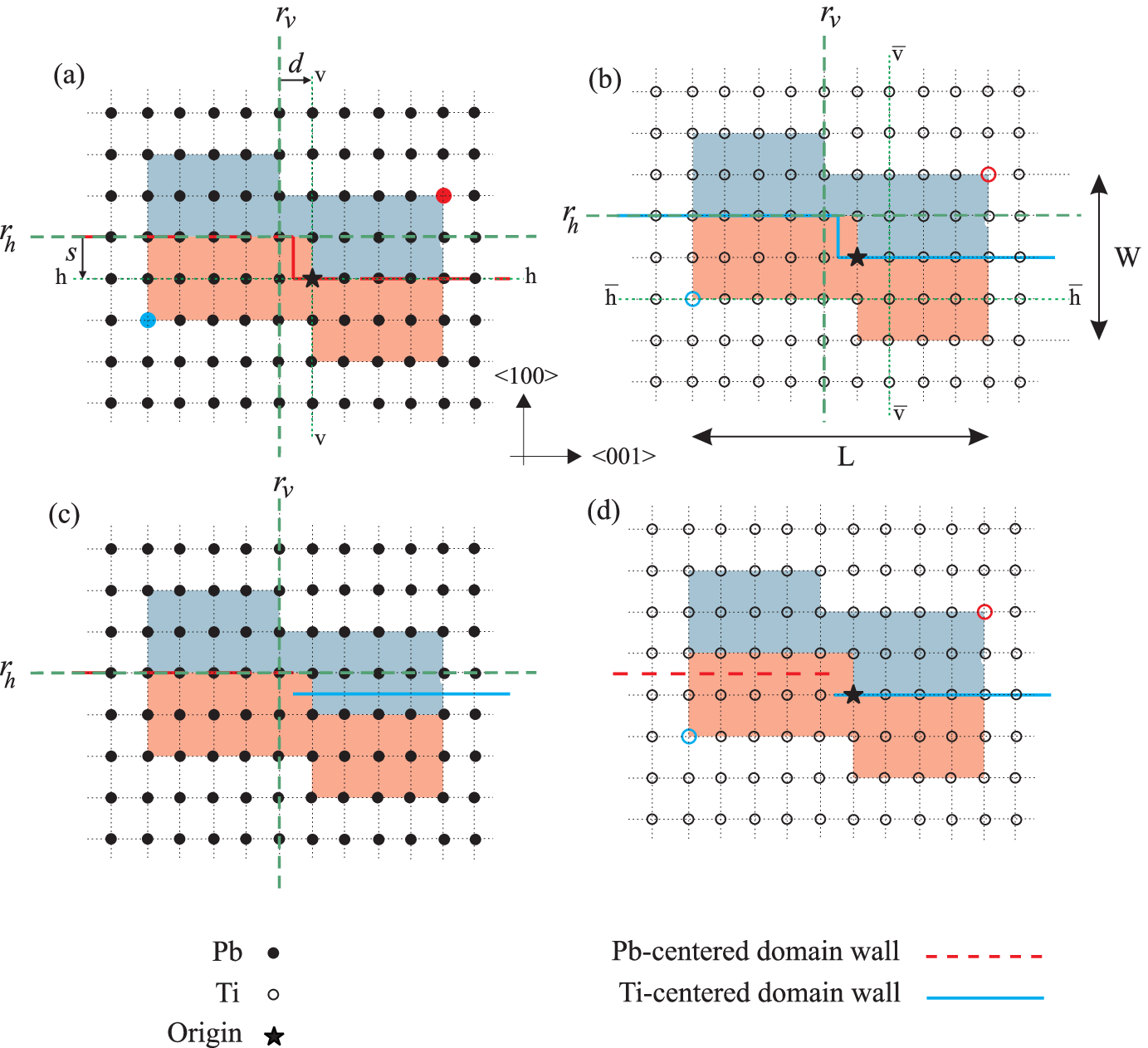}
\vspace*{-0.2in}
                 }}
\caption{A representative atomic layer for the initial configuration
of the three steps: (a) Pb cores in Pb-Pb step, (b) Ti cores in
Ti-Ti step, (c) Pb cores in Pb-Ti step, and (d) Ti cores in Pb-Ti
step. Note that planes h-h and v-v are sections that are used for a
better display of the variation of the distortion field in our
numerical examples. $s$ and $d$ denote the distances of sections h-h
and v-v from the reference planes $r_h$ and $r_v$, respectively.
$r_h$ and $r_v$ are parallel to $(100)$ and $(001)$-planes,
respectively. h-h and v-v sections in part (a) correspond to $s=a$
and $d=a$ and $\bar{\mathrm{h}}$-$\bar{\mathrm{h}}$ and
$\bar{\mathrm{v}}$-$\bar{\mathrm{v}}$ sections in part (b)
correspond to $s=2a$ and $d=2a$. The shaded regions denote the
computational box, which contains $W\times L$ unit cells and
different colors show the regions with opposite polarization inside
the computational box. The symbol $\bigstar$ in these figures denote the origin of the
coordinate system in each type of steps. The blue and red filled and
hollow circles denote the atoms whose displacements are used as
the displacements of the atoms located outside of the computational
box. \label{Steps}}
\end{figure*}

\section{Geometry of Steps} \label{sec:Geometry}

The geometry of the relaxed unit cell of tetragonal PbTiO$_3$ is
shown in Fig.\ref{UC}(a). The nonzero relative displacements in the
$\langle010\rangle$-direction between the center of the positive and
negative charges generate a polarization in the
$\langle010\rangle$-direction (we are using a shell potential). In
the $180^{\circ}$ domain walls, direction of polarization switches
across the domain wall. There are two types of $180^{\circ}$ domain
wall in PbTiO$_3$, namely, Ti-centered and Pb-centered domain walls.
Using the relaxed bulk configurations, it is possible to calculate
the atomic structure of both types \cite{YaOrBh2006b}.

Consider a domain wall parallel to a $(100)$-plane. By a step on the
domain wall we mean the region where the domain wall joins another
domain wall parallel to the first wall with an offset in the
$\langle100\rangle$-direction (see Figs.\ref{UC}(b) and (c)).\footnote{We assume that this offset is one or half a lattice spacing.} We
consider three different steps: Ti-Ti, Pb-Pb, and Pb-Ti.
Fig.\ref{Steps} shows the unrelaxed initial configuration for each
step. Note that assuming that the step is limited to a single unit
cell, i.e. if the two domain walls are one or half a lattice spacing
apart, there would be more than one possibility for the step
configuration. As an example, we plot two possibilities for Pb-Pb
step in Fig. \ref{UC}(c). In this figure, Case I shows a Pb-Pb step
in $(001)$ PbO-plane while Case II shows another Pb-Pb step in
$(001)$ TiO$_2$-plane. Note that there are still other possibilities
for Pb-Pb steps. We should emphasize that the configurations
shown in Fig. \ref{UC}(c) and Fig.\ref{Steps} are only the initial
configurations that we use as the starting point for finding the
final equilibrium configuration. We observe that as far as we
confine the step to a single unit cell, the anharmonic lattice
statics iterations converge to the same solution regardless of the
initial configuration of the step. Therefore, the exact choice of the
initial step configuration is not important in the final equilibrium
structure. We should also emphasize that we are analyzing a single step
on a single domain wall in an infinite crystal, i.e. no periodicity
assumptions are made. Note that in Fig.\ref{Steps}, domain walls
away from the step have polarization only along the
$\langle010\rangle$-direction. Note also that we assume a 2-D
symmetry reduction, which means that all the atoms with the same x and z coordinates (x, y, and z are coordinates along
 the $\langle100\rangle$, $\langle010\rangle$, and $\langle001\rangle$-directions, respectively) have the same displacements. Therefore, we partition the 3-D lattice
$\mathcal{L}$ as
$\mathcal{L}=\bigsqcup_{I}^{}\bigsqcup_{\alpha,\beta\in
\mathbbm{Z}}\mathcal{L}_{I\alpha\beta}$, where
$\mathcal{L}_{I\alpha\beta}$ and $\mathbbm{Z}$ are 1-D equivalence
classes parallel to the $\langle010\rangle$-direction and the set of
integers, respectively. See \cite{YaOrBh2006a,YavariAngoshtari2010}
for more details on the symmetry reduction.

\begin{figure*}[t]
\centerline{\hbox{\includegraphics[scale=.85,angle=0]{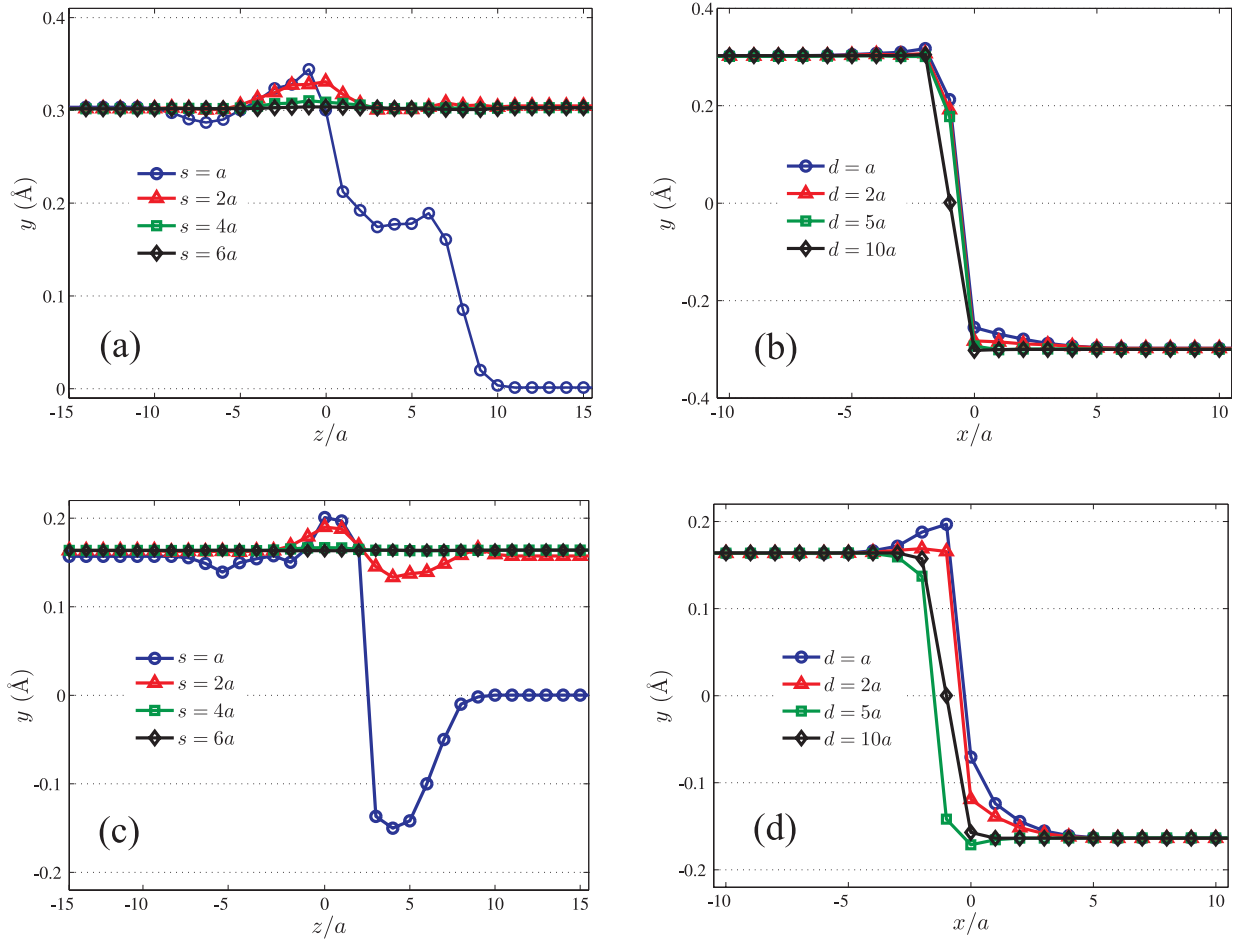}
\vspace*{-0.2in}
                 }}
\caption{The y-coordinates of atoms. (a) and (b) are Pb
cores in a Pb-Pb step, (c) and (d) are Ti cores in a Ti-Ti step.
Note that as it is shown in Fig.\ref{Steps}, $s$ and $d$ denote the
distances from the reference planes. \label{PPTT_y}}
\end{figure*}

\begin{figure*}[t]
\centerline{\hbox{\includegraphics[scale=.75,angle=0]{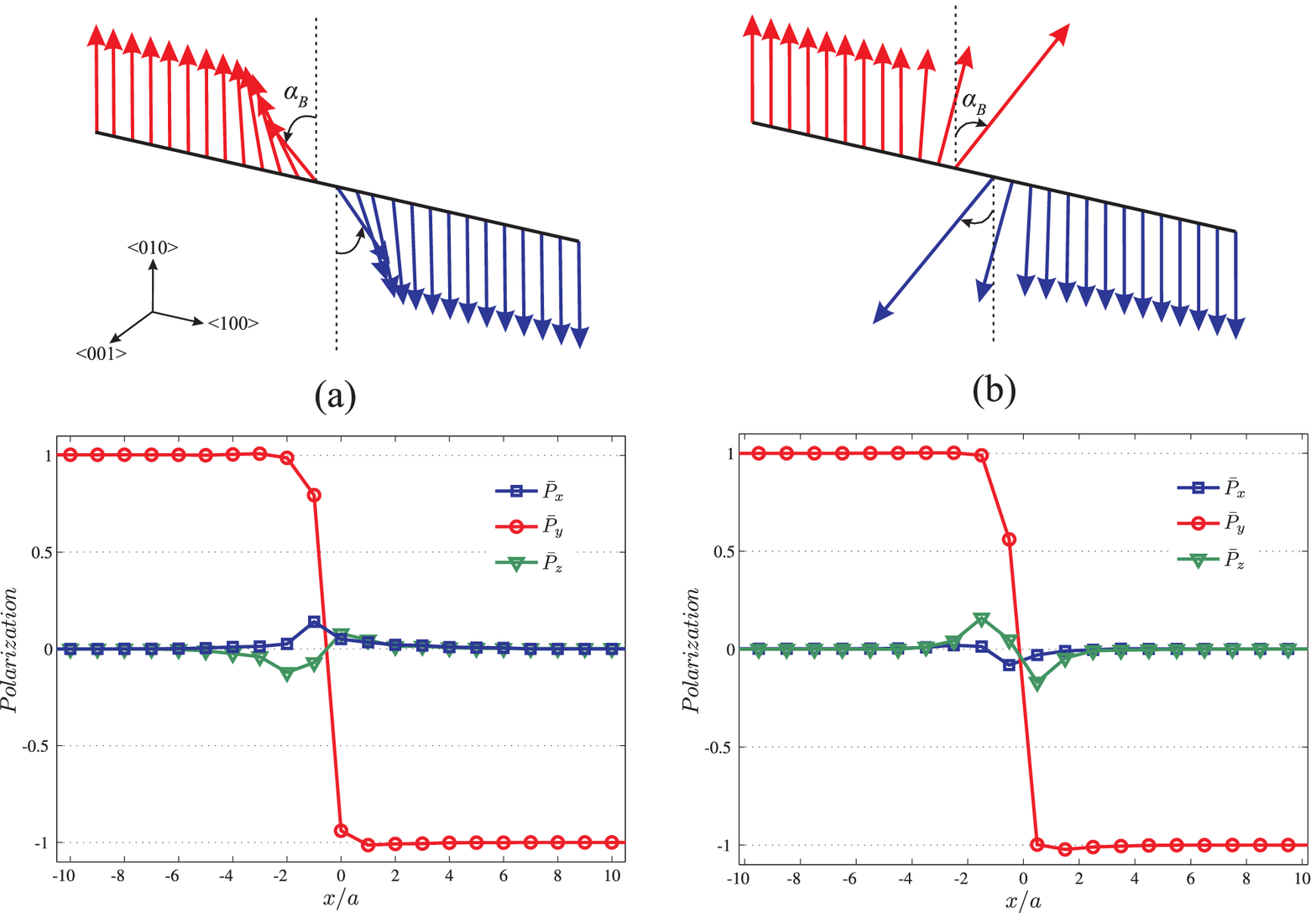}
\vspace*{-0.2in}
                 }}
\caption{The polarization vectors
$\overline{\mathbf{P}}=(\bar{P}_x,\bar{P}_y,\bar{P}_z)$ for the row
of unit cells on the section v-v with $d=a$ for: (a) Pb-Pb step and
(b) Ti-Ti step. Close to the step, polarization rotates out of the
$(001)$-plane with the Bloch angle $\alpha_B$. The polarization also
rotates inside the $(001)$-plane with the N\'{e}el angle $\alpha_N$.
Note that the Bloch and N\'{e}el components of the polarization
correspond to the components in $\langle001\rangle$-direction ($P_z$) and $\langle100\rangle$-direction ($P_x$), respectively.
\label{PolarPPTT}}
\end{figure*}

\section{Method of Calculation}\label{sec:Calculation}

We use the method of anharmonic lattice statics \cite{YaOrBh2006a} to calculate
the atomic structure of steps. We use a shell potential for
PbTiO$_3$ \citep{Asthagiri2006} to model the atomic interactions. In
this potential, each ion is represented by a core and a massless
shell. Let $\mathcal{L}$ denote the collection of cores and shells,
$i\in\mathcal{L}$ denote a core or a shell in $\mathcal{L}$, and
$\left\{\mathbf{x}^i \right\}_{i\in \mathcal{L}}$ represent the
current position of cores and shells. Then, the total static energy can
be written as
\begin{eqnarray}
    \mathcal{E}\left(\left\{\mathbf{x}^i \right\}_{i\in \mathcal{L}}
    \right)\!&=&\!\mathcal{E}_{\textrm{short}}\left(\left\{\mathbf{x}^i \right\}_{i\in \mathcal{L}}
    \right)
    + \mathcal{E}_{\textrm{long}}\left(\left\{\mathbf{x}^i \right\}_{i\in
    \mathcal{L}} \right)\nonumber \\
    &+&\mathcal{E}_{\textrm{core-shell}}\left(\left\{\mathbf{x}^i \right\}_{i\in \mathcal{L}}
    \right).
\end{eqnarray}
$\mathcal{E}_{\textrm{short}}\left(\left\{\mathbf{x}^i
\right\}_{i\in \mathcal{L}}\right)$ denotes short range
interactions, which are assumed to be only between Pb-O, Ti-O, and
O-O shells. The short range interactions are described by the
Rydberg potential of the form $(A+Br)\exp(-r/C)$, where A, B and C
are potential parameters and $r$ is the distance between interacting
elements. $\mathcal{E}_{\textrm{long}}\left(\left\{\mathbf{x}^i
\right\}_{i\in \mathcal{L}} \right)$ denotes the Coulombic
interactions between the core and shell of each ion with the cores
and shells of all of the other ions. Note that for calculating the
classical Coulombic potential and force, we use the damped Wolf
method \cite{Wolf99}. Finally,
$\mathcal{E}_{\textrm{core-shell}}\left(\left\{\mathbf{x}^i
\right\}_{i\in \mathcal{L}}\right)$ represents the interaction of
core and shell of an atom and is assumed to be an anharmonic spring
of the form $(1/2)k_2 r^2 +(1/24)k_4 r^4$, where $k_2$ and $k_4$ are
constants. All calculations are done for absolute zero temperature.
As is shown in Fig.\ref{UC}, at this temperature PbTiO$_3$ has a
tetragonal unit cell with lattice parameters $a=3.843~{\AA}$ and
$c=1.08a$ \cite{Asthagiri2006}. For more details on this notation
see \cite{YavariAngoshtari2010}.

For the relaxed configuration $\mathcal{B}=\left\{\mathbf{x}^i
\right\}_{i\in\mathcal{L}}\subset\mathbbm{R}^3$, static energy attains a
local minimum and hence we have
\begin{equation}\label{equilibrium}
    \frac{\partial \mathcal{E}}{\partial \mathbf{x}^i}=\mathbf{0}~~~~~~~\forall~i\in\mathcal{L}.
\end{equation}
To obtain the solution of the above optimization problem we use
the Newton method, which is based on a quadratic approximation near
the current configuration $\mathcal{B}^{k}$:
\begin{eqnarray}
    \mathcal{E}\left(\mathcal{B}^{k}+\tilde{\boldsymbol{\delta}}^k\right)&=&\mathcal{E}\left(\mathcal{B}^{k}\right)
    +\boldsymbol{\nabla}\mathcal{E}\left(\mathcal{B}^{k}\right)\cdot\tilde{\boldsymbol{\delta}}^k
    \nonumber \\
    &+&\frac{1}{2}(\tilde{\boldsymbol{\delta}}^k)^{\textsf{T}}\cdot\mathbf{H}\left(\mathcal{B}^{k}\right)
    \cdot\tilde{\boldsymbol{\delta}}^k+o\left(|\tilde{\boldsymbol{\delta}}^k|^2\right),
    \nonumber \\
\end{eqnarray}
where $\tilde{\boldsymbol{\delta}}^k=\mathcal{B}^{k+1}-\mathcal{B}^{k}$
and $\mathbf{H}$ is the Hessian matrix. In the Newton method:
\begin{equation}
    \tilde{\delta}^k=-\mathbf{H}^{-1}\left(\mathcal{B}^{k}\right)\cdot\boldsymbol{\nabla}\mathcal{E}\left(\mathcal{B}^{k}\right).
\end{equation}
Having $\tilde{\boldsymbol{\delta}}^k$, the next configuration is
calculated as:
$\mathcal{B}^{k+1}=\mathcal{B}^{k}+\tilde{\boldsymbol{\delta}}^k$.

As the size of the simulation box increases, the calculation of the
Hessian becomes inefficient and hence we use the quasi-Newton method.
In this method, instead of calculating the Hessian in each
iteration, one uses the Broyden-Fletcher-Goldfarb-Shanno (BFGS)
algorithm to approximate the inverse of the Hessian
\cite{PressTVF1989}. One starts with a positive-definite matrix and
uses the following BFGS algorithm to update the Hessian at each
iteration:
\begin{eqnarray}\label{BFGS}
    \mathbf{C}^{i+1}&=&\mathbf{C}^i + \frac{\tilde{\boldsymbol{\delta}}^k\otimes\tilde{\boldsymbol{\delta}}^k}{
    (\tilde{\boldsymbol{\delta}}^k)^{\textsf{T}}\cdot\mathbf{\Delta}}
    -\frac{\left(\mathbf{C}^{i}\cdot\mathbf{\Delta}\right)\otimes\left(\mathbf{C}^{i}\cdot\mathbf{\Delta}\right)}
          {\mathbf{\Delta}^{\textsf{T}}\cdot\mathbf{C}^{i}\cdot\mathbf{\Delta}} \nonumber \\
    &+&\left(\mathbf{\Delta}^{\textsf{T}}\cdot\mathbf{C}^{i}\cdot\mathbf{\Delta}\right)\mathbf{u}\otimes\mathbf{u},
\end{eqnarray}
where $\mathbf{C}^{i}=\left(\mathbf{H}^i\right)^{-1}$,
$\mathbf{\Delta}=\boldsymbol{\nabla}\mathcal{E}^{i+1}-\boldsymbol{\nabla}\mathcal{E}^{i}$, and
\begin{eqnarray}
  \mathbf{u}=\frac{\tilde{\boldsymbol{\delta}}^k}{(\tilde{\boldsymbol{\delta}}^k)^{\textsf{T}}\cdot\mathbf{\Delta}}-
  \frac{\mathbf{C}^{i}\cdot\mathbf{\Delta}}{\mathbf{\Delta}^{\textsf{T}}\cdot\mathbf{C}^{i}\cdot\mathbf{\Delta}}.
\end{eqnarray}
Calculating $\mathbf{C}^{i+1}$, one then should use
$\mathbf{C}^{i+1}$ instead of $\mathbf{H}^{-1}$ to update the
current configuration for the next configuration
$\mathcal{B}^{k+1}=\mathcal{B}^{k}+\tilde{\boldsymbol{\delta}}^k$.
If $\mathbf{C}^{i+1}$ is a poor approximation, then one may need to
perform a linear search to refine $\mathcal{B}^{k+1}$ before
starting the next iteration.

As the initial configuration for each step we start with two half
lattices with the proper offset in the x-direction. The atomic
configuration of each half lattice is the same as the atomic
configuration in a perfect $180^{\circ}$ domain wall. To remove the
rigid body translation of the lattice, we fix the core of an atom in
our computational box and fully relax the other atoms. Hence, we
have $30W\times (L-3)$ variables in our calculations, where $W$ and
$L$ are specified in Fig.\ref{Steps}. To consider the effect of the
atoms outside the computational box, we impose rigid body
translations to these atoms as the boundary conditions, i.e., we
rigidly move the atoms outside the computational box such that they
keep the perfect $180^{\circ}$ domain wall configuration. To this
end, we rigidly move all of the atoms outside the computational box
in the positive (negative) direction of the z-axis with the
displacements equal to the displacement of the first (last) atom of
the first (last) row of the representative layer of atoms. This is
marked with the red (blue) circle in Fig.\ref{Steps}(a). The
displacements of the atoms outside the computational box in the
positive (negative) direction of the x-axis are equal to the
displacements of the atoms in the first (last) row of the
computational box that is located on the same column. We choose
$W=20$ and $L=30$ as we see larger values will not affect the
results. In all our calculations we assume force tolerance of $0.05~
\textrm{eV/{\AA}}$ and observe that our solutions converge slowly
after about $800$ to $1000$ iterations depending on the step type.
Our calculations also show that using a smaller force tolerance
of $0.005~ \textrm{eV/{\AA}}$ would change the results by less than
$0.1\%$. This justifies the above choice of the force tolerance.

\begin{figure*}[t]
\centerline{\hbox{\includegraphics[scale=0.9,angle=0]{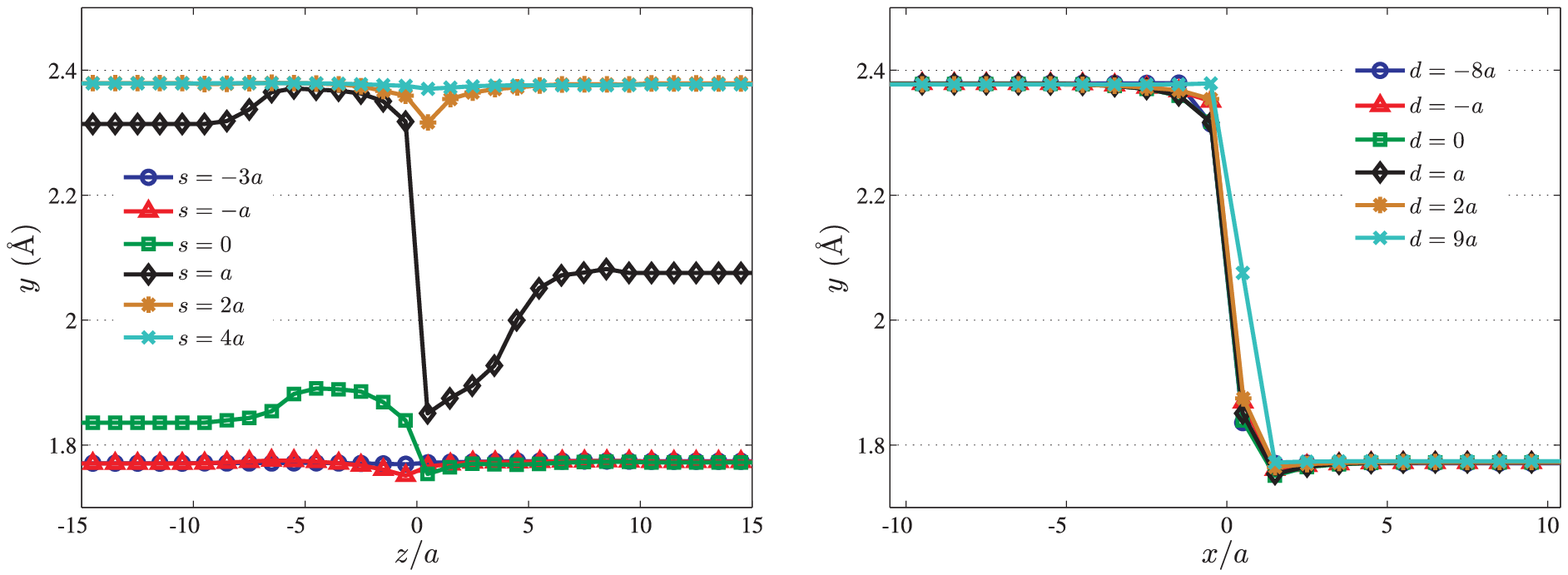}
\vspace*{-0.2in}
                 }}
\caption{The y-coordinates of Pb cores in a Pb-Ti step. Note that as it
is shown in Fig.\ref{Steps}, $s$ and $d$ denote the distances from
the reference planes.  \label{PbTi_y}}
\end{figure*}

\begin{figure}[t]
\begin{center}
\includegraphics[scale=0.65,angle=0]{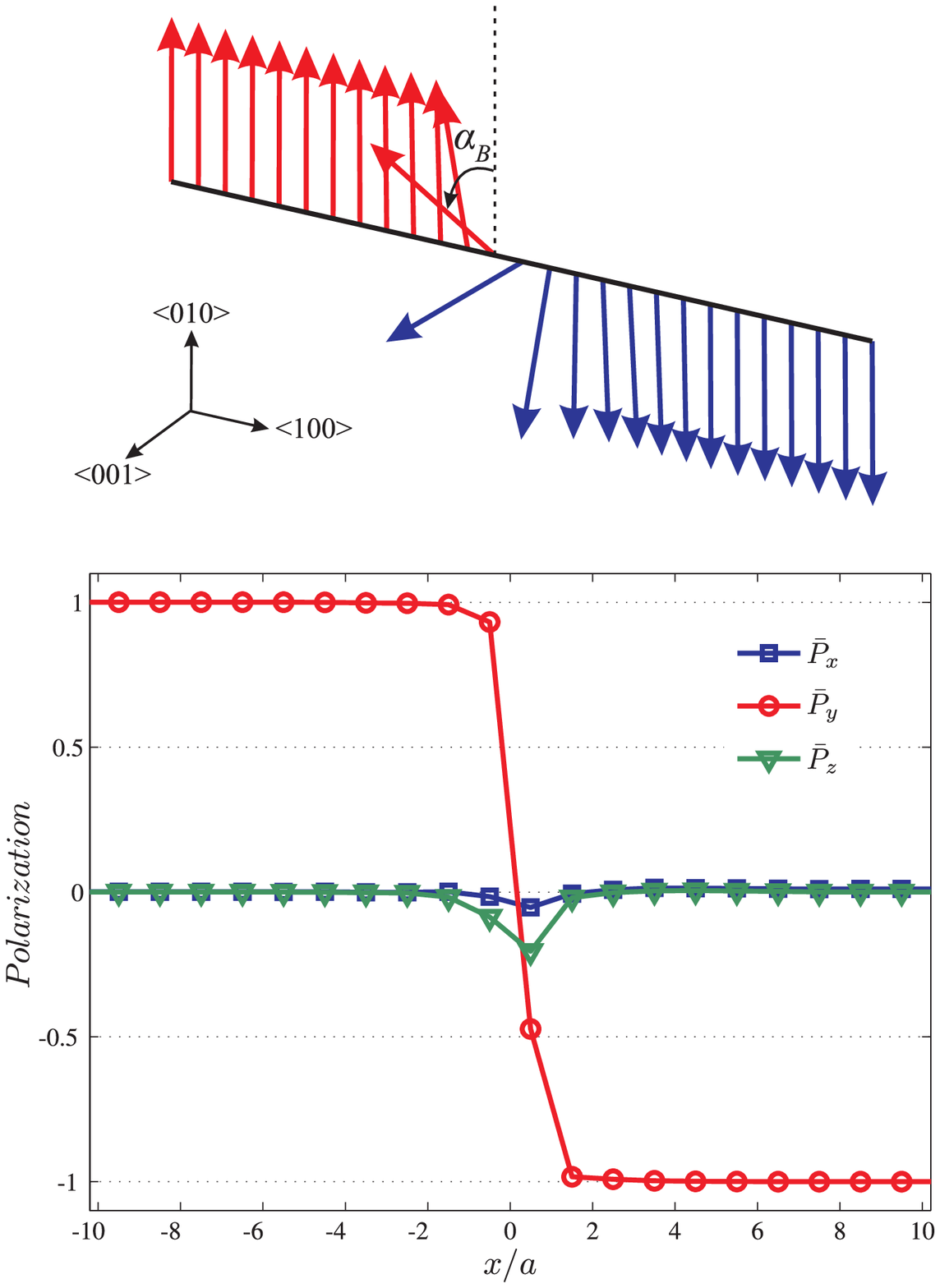}
\end{center}\vspace*{-0.2in}
\caption{The polarization vectors
$\overline{\mathbf{P}}=(\bar{P}_x,\bar{P}_y,\bar{P}_z)$ for the row
of unit cells on the section v-v with $d=a$ for a Pb-Ti step. Close
to the step, polarization rotates out of the $(001)$-plane with the
Bloch angle $\alpha_B$. The polarization also rotates inside the
$(001)$-plane with the N\'{e}el angle $\alpha_N$. Note that the
Bloch and N\'{e}el components of the polarization correspond to the
components in $\langle001\rangle$-direction ($P_z$) and
$\langle100\rangle$-direction ($P_x$), respectively.}
\label{PolarPT}
\end{figure}

\section{Numerical Results }\label{sec:Results}

In this section we present our numerical results for the three different steps
as follows. As we mentioned earlier, x, y, and z are coordinates
along the $\langle100\rangle$, $\langle010\rangle$, and
$\langle001\rangle$-directions, respectively, and the origin of the
coordinate system for each step is specified in Fig.\ref{Steps}.

\textbf{Pb-Pb step:} The atomic configuration of Pb-cores in a Pb-Pb
step are shown in Fig.\ref{PPTT_y}(a) and (b). For a clearer
presentation of the atomic configuration, we have plotted the
y-coordinates of Pb-cores for different sections v-v and h-h (see
Fig.\ref{Steps}). In Fig.\ref{PPTT_y}, $s$ and $d$ denote the
distances of the h-h and v-v sections from the reference planes,
respectively. Note that as is shown in Fig.\ref{Steps}, the
reference plane for v-v sections ($r_v$) is parallel to the
$(001)$-plane and the reference plane for h-h sections ($r_h$) is
parallel to the $(100)$-plane. As it can be seen, atomic distortions
in the Pb-Pb step are localized, i.e, they are confined to a
$8a\times 20a$ box in the $(010)$-plane. Atomic distortions in the
$\langle001\rangle$-direction are less localized compared to those
in the $\langle100\rangle$-direction. We observe that the step
thickens the domain walls; the width of the domain wall near the
step is about three times that of the prefect Pb-centered domain
wall. Note that domain wall thickness cannot be defined uniquely
very much like boundary layer thickness in fluid mechanics. Here,
domain wall thickness is by definition the region that is affected
by the domain wall, i.e. those layers of atoms that are distorted.
One can use definitions like the $99\%$-thickness in fluid mechanics
and define the domain wall thickness as the length of the region
that has $99\%$ of the far field rigid translation displacement.
What is important here is that no matter what definition is chosen,
domain wall ``thickness" increases by the presence of steps. Note
that due to the symmetry of the Pb-Pb step, atomic configuration for
negative values of $s$ and $d$ will have the same behavior.  Also
note that the y-components of the atoms on the section $s=a$ in
Fig.\ref{PPTT_y}(a) are not symmetric because of the way we define this
section (see Fig.\ref{Steps}). As the coordinates of cores and
shells are close to each other, we only plot the results for cores.
Also because other types of atoms display a similar behavior, we do
not plot their coordinates here.

We follow Meyer and Vanderbilt \cite{MeyerVanderbilt2001} to
calculate the polarization profile for each step. The polarization of
unit cell $i$ is calculated as
\begin{eqnarray}\label{Polarization}
  \mathbf{P}_{i}=\frac{e}{\Omega_c}\sum_{j}w_{j}Z_{j}^{*}\mathbf{u}^{i}_{j},
\end{eqnarray}
where $e$ is the electron charge, $\Omega_c$ is the volume of the
unit cell, $Z_{j}^{*}$ is the Born effective charge tensor of the
cubic PbTiO$_{3}$ bulk, and $\mathbf{u}^{i}_{j}$ denotes the
displacement of the $j$-th atom of the unit cell $i$ from the ideal
lattice site. $w_j$ denotes the weight for atom $j$. For example,
for a Ti-centered unit cell we have $w_{Ti}=1$, $w_{O}=1/2$, and
$w_{Pb}=1/8$. We have plotted the polarization
$\overline{\mathbf{P}}_{i}$ of the rows of unit cells on the section
v-v with $d=a$ (see Fig.\ref{Steps}) in a Pb-Pb step in
Fig.\ref{PolarPPTT}(a), where $\overline{\mathbf{P}}_{i}=
\mathbf{P}_{i}/|\mathbf{P}_{b}|$, with $|\mathbf{P}_{b}|$ denoting
the norm of the polarization of the bulk. We obtain the bulk
polarization of $80.1~\mu C~cm^{-2}$, which is close to the
published values $81.0~\mu C~cm^{-2}$ \cite{BeharaHinojosa2008} and
$81.2~\mu C~cm^{-2}$ \cite{MeyerVanderbilt2001}. We observe that
near the step, the domain wall has a mixed Bloch-N\'{e}el character.
Denoting the polarization components by
$\overline{\mathbf{P}}=(\bar{P}_x,\bar{P}_y,\bar{P}_z)$, where
$\bar{P}_x$, $\bar{P}_y$, and $\bar{P}_z$ are polarization
components in $\langle100\rangle$, $\langle010\rangle$, and
$\langle001\rangle$-directions, respectively, we observe that the
polarization vector rotates out of the $(001)$-plane with the Bloch
angle $\alpha_{B}=\tan^{-1}\left(\bar{P}_z/\bar{P}_y\right)$ (see
Fig.\ref{PolarPPTT}). The maximum rotation angle $\alpha_B$ for
Pb-centered domain wall is $\alpha_B \simeq 7.0^{\circ}$. Also the
polarization rotates in the $(001)$-plane with the N\'{e}el angle
$\alpha_{N}=\tan^{-1}\left(\bar{P}_x/\bar{P}_y\right)$. The maximum
value of $\alpha_{N}$ for Pb-centered wall is $\alpha_N \simeq
9.9^{\circ}$ (compare this with $\alpha_{N}=1.43^{\circ}$ for the
perfect domain wall \cite{Lee2009}). The maximum value of the
polarization in the $\langle100\rangle$ and $\langle001\rangle$-directions are about $13.9\%$ and $12.2\%$ of the bulk polarization,
respectively.

Finally, we calculate the energy of the Pb-Pb step, $E_{Pb-Pb}$.
Similar to the domain wall energy, we define the step energy to be
the difference in energies of the unit cells inside the
computational box that are located on the domain wall with the step
and bulk energy of the same number of unit cells, divided by the
total area of domain wall in the system. This way we obtain the Pb-Pb
step energy to be $157~mJm^{-2}$, which is greater than the
Pb-centered domain wall energy that is $132~mJm^{-2}$
\cite{MeyerVanderbilt2001}.

\textbf{Ti-Ti step:} Figs.\ref{PPTT_y}(c) and (d) depict the
y-coordinates of Ti cores in a Ti-Ti step for different sections v-v
and h-h. Again because of symmetry, we plot the results only for
positive values of $s$ and $d$ and also similar to the Pb-Pb step, by definition of the section $s=a$, the y-components of the
atoms on this section in Fig.\ref{PPTT_y}(c) are not symmetric.
Similar to the Pb-Pb step, we observe that the Ti-Ti step is
localized, i.e., atomic distortions are confined to a $9a\times18a$
box in the $(010)$-plane. Again we observe that the step thickens
the Ti-centered domain wall; the thickness of the defective wall is
about three times that of the perfect domain wall.

As is shown in Fig.\ref{PolarPPTT}(b), polarization has a mixed
Bloch-N\'{e}el character near the step. For the Ti-Ti step, the
maximum value of the Bloch and N\'{e}el rotation angles are
$\alpha_B \simeq 9.5^{\circ}$ and $\alpha_N \simeq 8.1^{\circ}$
(compare this with  $\alpha_N = 1.0^{\circ}$ in the perfect domain
wall \cite{Lee2009}), respectively. The maximum value of the
polarization in the $\langle100\rangle$ and $\langle001\rangle$-directions are about $8.1\%$ and $16.8\%$ of the bulk polarization,
respectively. The energy of the Ti-Ti step is
$E_{Ti-Ti}=172~mJm^{-2}$, which is larger than the energy of the
Pb-Pb step and the energy of the Ti-centered domain wall, which is
$169~mJm^{-2}$ \cite{MeyerVanderbilt2001}. This is consistent with
the fact that Ti-centered $180^{\circ}$ domain walls have a greater
static energy than Pb-centered domain walls
\cite{MeyerVanderbilt2001,AngYa2010}.

\textbf{Pb-Ti step:} We have plotted the y-coordinates of Pb cores
in a Pb-Ti step in Fig.\ref{PbTi_y} for different v-v and h-h
sections. Note that because Pb-Ti steps are not symmetric, we have
plotted the results for both positive and negative values of $s$ and
$d$. Also since other types of cores and shells have a similar
behavior, we do not plot their coordinates here. We observe that
similar to the other two steps, the Pb-Ti step causes local
distortions that are confined to a $6a\times 16a$ box in
$(010)$-plane. The step broadens the domain wall; the defective
domain wall thickness is twice that of the perfect domain wall. Note that the Pb-centered and Ti-centered domain walls for this step are half a lattice spacing apart and this may explain the weaker thickening effect of the Pb-Ti step.

As Fig.\ref{PolarPT} shows polarization distribution has a mixed
Bloch-N\'{e}el character near the Pb-Ti step but the Bloch character is
more dominant. The Polarization profile is plotted for the row of
unit cells on the section v-v with $d=a$, which is located in the
Ti-centered part of the step.  For the Pb-Ti step, the maximum value
of the Bloch and N\'{e}el rotations are $\alpha_B \simeq
23.0^{\circ}$ and $\alpha_N \simeq 5.9^{\circ}$, respectively. The
maximum value of the polarization in the $\langle100\rangle$ and
$\langle001\rangle$-directions are about $5.3\%$ and $20.5\%$ of the
bulk polarization, respectively. The energy of the Pb-Ti step is
$E_{Pb-Ti}=165~mJm^{-2}$. It is seen that
$E_{Pb-Pb}<E_{Pb-Ti}<E_{Ti-Ti}$.

\section{Concluding Remarks}\label{sec:Concluding}

In this work we obtained the atomic structure of three different
types of steps on $180^{\circ}$ domain walls in PbTiO$_3$ using the
method of anharmonic lattice statics. We observe that these steps
cause local atomic distortions that are confined to a box with
dimensions of a few lattice spacings. All the three steps have a
broadening effect on the domain wall thickness. Pb-Ti steps have
a less broadening effect compared to the other two steps.

We also observe that steps on $180^{\circ}$
domain walls can cause the polarization profile to have a mixed
Bloch-N\'{e}el character. The Bloch character is more dominant in
Ti-Ti and Pb-Ti steps. Finally, we observe that the Pb-Pb step has
a lower static energy than the other two steps.

\acknowledgments Study of steps on ferroelectric domain walls was
suggested to the second author by Professors Michael Ortiz and
Kaushik Bhattacharya. We thank an anonymous referee for
useful comments that improved the presentation of our work.


\begin{references}
\bibitem{Kalinin2010} S. V. Kalinin, A. N. Morozovska, L. Q. Chen and B. J. Rodriguez,
              {\it Rep. Prog. Phys.}, {\bf 73} , 056502 (2010).

\bibitem{DawberRabeScott2005} M. Dawber, K. M. Rabe and J. F. Scott,
              {\it Rev. of Modern Phys.}, {\bf 77} , 1083 (2005).

\bibitem{MeyerVanderbilt2001} B. Meyer and D. Vanderbilt
             {\it Phys. Rev. B}, {\bf65}, 1 (2001).

\bibitem{Padilla1996} J.  Padilla, W. Zhong and D. Vanderbilt
             {\it Phys. Rev. B}, {\bf53}, R5969 (1996).

\bibitem{YaOrBh2006b} A. Yavari, M. Ortiz and K. Bhattacharya,
              {\it Philos. Mag.}, {\bf 87} , 3997 (2007).

\bibitem{AngYa2010} A. Angoshtari and A. Yavari,
             {\it EPL}, {\bf90}, 27007 (2010).

\bibitem{HlinkaMarton2006} J. Hlinka and P. Marton,
             {\it Phys. Rev. B}, {\bf74}, 104104 (2006).

\bibitem{Iwata2003} M. Iwata, K. Katsuraya, I. Suzuki, M. Maeda, N. Yasuda and Y. K. Ishibashi,
             {\it Jpn. J. Appl. Phys.}, {\bf42}, 6201 (2003).

\bibitem{Lehnen2000} P. Lehnen, J. Dec and W. Kleemann,
             {\it J. Phys. D: Appl. Phys.}, {\bf33}, 1932 (2000).

\bibitem{Shilo2004} D. Shilo, G. Ravichandran and K. Bhattacharya,
             {\it Nature Mater.}, {\bf3}, 453 (2004).

\bibitem{Lee2005} W. T. Lee, E. K. H. Salje and U. Bismayer,
             {\it Phys. Rev. B}, {\bf72}, 104116 (2005).

\bibitem{AngYavari2010} A. Angoshtari and A. Yavari,
             {\it Comput. Mater. Sci.}, {\bf48}, 258 (2010).

\bibitem{Lee2009} D. Lee, R. K. Behera, P. Wu, H. Xu, Y. L. Li, S. B. Sinnott, S. R. Phillpot, L. Q. Chen and V. Gopalan,
             {\it Phys. Rev. B}, {\bf80}, 060102(R) (2009).

\bibitem{NaumovBellaicheFu2004} I. I. Naumov, L. Bellaiche and H. X. Fu,
             {\it Nature}, {\bf432}, 737 (2004).

\bibitem{Prosandeev2008} S. Prosandeev, I. Ponomareva, I. Naumov, I. Kornev and L. Bellaiche,
             {\it J. Phys.: Condens. Matter}, {\bf20}, 193201 (2008).

\bibitem{Nettleton1967} R. E. Nettleton,
             {\it J. Phys. Soc. Jpn.}, {\bf22}, 1375 (1967).

\bibitem{Shur1990} V. Y. Shur, A. L. Gruverman and E. L. Rumyantsev,
             {\it Ferroelec.}, {\bf111}, 123 (1990).

\bibitem{ShinGrinberg2007} Y. H. Shin, I. Grinberg, I. W. Chen and A. M. Rappe,
             {\it Nature}, {\bf449}, 881 (2007).

\bibitem{YaOrBh2006a} A. Yavari, M. Ortiz and K. Bhattacharya,
              {\it J. Elasticity}, {\bf 86}, 41 (2007).

\bibitem{YavariAngoshtari2010} A. Yavari and A. Angoshtari,
              {\it Inter. J. Solids Struct.}, {\bf 47}, 1807 (2010).

\bibitem{Asthagiri2006} A. Asthagiri, Z. Wu, N. Choudhury and R. E. Cohen,
              {\it Ferroelec.}, {\bf333} , 69 (2006).

\bibitem{BeharaHinojosa2008} R. K. Behera, B. B. Hinojosa, S. B. Sinnott, A. Asthagiri \and S. R. Phillpot,
              {\it J. Phys.: Condens. Matter}, {\bf20}, 395004 (2008).

\bibitem{Wolf99} D. P. Wolf, P. Keblinski, S. R. Phillpot and J. Eggebrecht,
              {\it J. Chem. Phys.}, {\bf110}, 8254 (1999).

\bibitem{PressTVF1989} W. H. Press, S. A. Teukolsky, W. T. Vetterling, and B. P. Flannery, {\it Numerical recipes: the art of
              scientific computing}
             (Cambridge University Press, Cambridge, 1989).



















%
%
%

%

%

%
%
%
%
%
%
%



%
%
%
%

%

%
%
%
%
%
%
%

%

%
%

%


\end{references}
\end{document}